\documentclass{PoS}
\newcommand\beq{\begin{eqnarray}}
\newcommand\eeq{\end{eqnarray}}
\newcommand\nn{\nonumber}
\newcommand {\bmp}{\begin{minipage}}
\newcommand {\emp}{\end{minipage}}
\newcommand \reffig {Fig.~\ref}
\newcommand \refeq[1]{Eq.~(\ref{#1})}

\title{Parity doubling in two-flavor SU(2) 
at high temperature}

\ShortTitle{Parity Doubling in two-flavor SU(2) 
at high temperature}

\author{\speaker{Jong-Wan Lee}
\thanks{Currently at Extreme Physics Institute, Pusan National 
	University, Busan 46241, Korea.}\\
        E-mail: \email{j.w.lee@swansea.ac.uk}}

\author{Biagio Lucini\\
        E-mail: \email{b.lucini@swansea.ac.uk}}

\author{Maurizio Piai\\

        E-mail: \email{m.piai@swansea.ac.uk}\\

	Department of Physics, College of Science, Swansea University, Singleton Park, SA2 8PP, Swansea, Wales, UK}

\abstract{
We study the mass spectrum of mesons at high temperature in $SU(2)$ gauge theory 
with two flavors of  Dirac fundamental fermions. 
Numerical simulations are carried out on anisotropic lattices 
using Wilson fermions, with lattice parameters  tuned 
so that Euclidean symmetry is restored at low energy. 
We determine the pseudo-critical temperature $T_c$ using renormalized Polyakov loops. 
We calculate temporal and spatial meson correlation functions 
across $T_c$, and observe a clear sign of parity doublings above $T_c$ 
in both vector and scalar channels. 
The degeneracy between parity partners in the spectrum
indicates that  the enhanced $SU(4)\times U(1)_A$ global symmetry of the model is restored at high temperature.}

\FullConference{34th annual International Symposium on Lattice Field Theory\\
		24-30 July 2016\\
		University of Southampton, UK}

\begin{document}

\section{Introduction}
\label{sec:introduction}

Two-color QCD, the gauge theory with $SU(2)$ gauge group, 
has been studied extensively by the lattice community 
as a laboratory to investigate from first principles the phase diagram in the $T$-$\mu$ plane
(see~\cite{QCD2} and references therein). 
While it shares some qualitative  similarities with the non-perturbative features of QCD, 
such as chiral symmetry breaking and confinement, 
the theory with an even number of flavors is not affected by the sign problem,
and its behavior at finite chemical potential $\mu$ can be studied 
using Monte-Carlo techniques.

In addition, the theory with two flavors of fundamental Dirac fermions 
has received considerable attention in its own terms: 
it is a template for new strong dynamics 
in the context of physics beyond the standard model (BSM). 
Depending on how the electroweak gauge group is embedded in the global symmetry, 
it can provide a microscopic realization of technicolor, composite Higgs, or admixtures of the two~\cite{Sannino}. 
In addition, a dark matter candidate can also appear naturally 
at low energy~\cite{Pica2012}. 
The dynamical features of utmost importance  for model building are related to the
underlying pattern of (spontaneous) symmetry breaking,
in particular in reference to 
 the appearance of  Nambu-Goldstone (NG) modes. 

At zero temperature, the meson spectrum~\cite{Pica2012} shows 
the appearance of five NG bosons, due to the breaking of the 
enhanced $SU(4)$ global symmetry to $Sp(4)$.
 In this work we investigate symmetry restoration at deconfinement (and for $\mu=0$),
which manifests itself in the degeneracy of masses
between mesons related by symmetry transformations.

We perform unquenched lattice simulations using $N_f=2$ Dirac-Wilson fermions 
on anisotropic lattices. 
Adapting anti-periodic boundary conditions for the temporal fermion action, 
the temperature $T$ is controlled by $N_t$ as $T=\frac{1}{N_t a_t}$,
where $N_t$ is the number of sites in the temporal direction, while $a_t$ is the lattice spacing. 
We determine the pseudo-critical temperature $T_c$ using the renomalized Polyakov loops. 
We  then calculate the two-point temporal and spatial correlation functions 
of flavored pseudo-scalar (PS), scalar (S), vector (V), and axial-vector (AV) mesons across $T_c$ 
and  measure the screening masses. 
For complete details of this study see~\cite{Lee}.

\section{The model}
\label{model}

The continuum action of the model in  Minkowski space is 
\beq
\mathcal{L}=
i\,\overline{Q^i}_{\,a}\,\gamma^{\mu}\,(D_{\mu}Q^i)^a\,-\,m\,\overline{Q^i}_{\,a}Q^{i\,a}\,
-\,\frac{1}{2}\textrm{Tr}\, F_{\mu\nu} F^{\mu\nu}\,,
\eeq
where summations are understood over the flavor index $i=1, 2$ and the color index $a=1, 2$. 
The covariant derivative is given by
\beq
(D_\mu Q^i)^a=\partial_{\mu}Q^{i\,a}+i g V^A_{\mu}(T^A)^a_{\,\,b}Q^{i\,b}\,,
\eeq
while the field strength tensors are $F_{\mu\nu}=\partial_\mu V_\nu-\partial_\nu V_\mu 
+ig[V_\mu,V_\nu]$.  
Each (massive) Dirac fermion $Q^{i\,a}$ can be rewritten as two $2$-component spinors $q^{i\, a}$ and $q^{i+2\,a}$. 
Because of the pseudo-real nature of $SU(2)$, in the massless limit the global symmetry of the action is enhanced to 
$U(1)_A\times SU(4)$. 
Let us focus on the breaking of $SU(4)$ symmetry by the mass term $\mathcal{L}_m=-m\overline{Q^i}_{\,a} Q^{i\,a}$,
and rewrite it in terms of $q^{i\,a}$~\cite{Lee}: 
\beq
\mathcal{L}_m
=\,-\frac{1}{2}m\epsilon_{ab}q^{n\,a\,T}(-i\tau^2) q^{m\,b}\,\Omega_{nm}\,+\,{\rm h. c.}\,,
\eeq
where the Pauli matrix $\tau^2$ acts on the spinor space. 
The antisymmetric matrix $\Omega$ is defined by
\beq
\Omega=
\left(\begin{array}{cc}
0 & 1_{2\times 2} \cr
-1_{2\times 2} & 0\cr
\end{array}\right)\,.
\label{eq:omega}
\eeq

The kinetic term of the action is invariant under $SU(4)$ transformations, 
and in particular under infinitesimal transformations 
$q\rightarrow \left(1+i\sum_A^{15} \alpha^A T^A\right)q$, with
$T^A$ the generators of $SU(4)$.
${\cal L}_m$ enjoys the same formal invariance, provided the only non-vanishing $\alpha^A$ are those for which
$\Omega T^A+(T^{A})^{T} \Omega=0$, which is the definition of $Sp(4)$.
A non-zero vacuum expectation value $\langle\bar{Q}Q\rangle\propto \Omega$
spontaneously breaks the symmetry in the same way. 
The consequence of this symmetry breaking is the appearance of the
five aforementioned NG bosons.

As with QCD,
the mass splittings in the meson spectra are controlled by symmetry breaking. 
While the mass of the pseudo-scalars $\pi^A$ ($J^P=0^-$) is protected by global symmetry, 
the $a_0^A$ ($0^+$) scalars are expected to have mass of order of the symmetry breaking scale. 
Furthermore, $U(1)_A$ symmetry is anomalously broken.
Hence, the restoration of $SU(4)\times U(1)_A$ global symmetries
results in the degeneracy of $\pi$ with $a_0$ masses.
The long distance dynamics of the $10$ spin-$1$ vector $\rho$ ($1^-$) and $5$ axial-vector $a_1$ ($1^+$) mesons can be described  
by generalizing hidden local symmetry. 
Their mass difference 
arises because of the $SU(4)\rightarrow Sp(4)$ symmetry breaking.
In other words, mass-degenerate $\rho$ and $a_1$ mesons isolate the symmetry restoration of $SU(4)$. 

\section{Lattice simulations}
\label{lattice}

Numerical simulations are carried out using the Wilson gauge action with two 
mass-degenerate Wilson-Dirac fundamental fermions on an anisotropic lattice, 
\beq
S_g=\frac{\beta}{\xi^0_g}
\left[
\sum_i (\xi^0_g)^2 \left(1-\frac{1}{2}\textrm{Re}~\textrm{tr}P_{0i}\right)
+\sum_{i<j}\left(1-\frac{1}{2}\textrm{Re}~\textrm{tr}P_{ij}\right)
\right],
\eeq
where $P$ denotes the plaquette and
\beq
S_f&=&a_s^3\sum_{\bf n}\bar{\psi}_{\bf n}
\left[
-\frac{1}{2\xi^0_f}\sum_j\left((1-\gamma_j)U_{{\bf n}, j}\psi_{{\bf n}+\hat{j}}+
(1+\gamma_j)U_{{\bf n}-\hat{j},j}^\dagger\psi_{{\bf n}-\hat{j}}\right)
\right.\nn \\
&&\left.
-\frac{1}{2}\left((1-\gamma_0)U_{{\bf n},0}\psi_{{\bf n}+\hat{0}}+
(1+\gamma_0)U_{{\bf n}-\hat{0},0}^\dagger\psi_{{\bf n}-\hat{0}}\right)
+\left(a_t m_0+1+\frac{3}{\xi_f^0}\right)\psi_{\bf n}\right],
\eeq
where $\beta=4/g^2$ and $m_0$ are the lattice bare gauge coupling 
and bare fermion mass, respectively. 
We have two additional bare parameters, $\xi_g^0$ and $\xi_f^0$, 
representing the gauge and fermion anisotropies. 
To recover Euclidean symmetry at low energy, 
these bare parameters are  tuned 
so that  the renormalized parameters satisfy $\xi_g=\xi_f=\xi$. 

Configurations are generated using the Hybrid Monte Carlo (HMC) algorithms 
with the second order Omelyan integrator for Molecular Dynamics (MD) evolution,
where different lengths of MD time steps $\delta \tau_\mu$ are used for gauge and fermion actions 
such that acceptance by Metropolis test is in the range of $75-85\%$.
The simulation codes are developed from the HiRep code~\cite{HiRep} modified by implementing the gauge and fermion
anisotropies described above.
To optimise the acceptance, we also treat the variances of temporal and spacial 
conjugate momenta differently by introducing a new tunable parameter. 
Without changing the validity of the method, such a setup is useful for the anisotropic lattice 
where the temporal and spacial MD forces are different \cite{MSU}. 
Thermalization and autocorrelation times are estimated by monitoring the average plaquette expectation values.
The statistical errors are obtained using the standard bootstrapping technique.
Throughout this study, the gauge coupling is fixed by $\beta=2.0$. 
\subsection{Anisotropic lattice}
\label{anisotropic_lattice}

To tune the lattice parameters we perform zero temperature calculations on 
a $128\times 12^3$ lattice with periodic boundary conditions
in all directions of both link variables and fermion fields.
Twelve ensembles are generated using various combinations of $(m_0,\,\xi_g,\,\xi_f)$ 
over the ranges $(-0.195\sim-0.215,\,4.5\sim5.1,\,4.5\sim4.9)$. 
For each ensemble $N_{conf}=138-300$ configurations are accumulated after $200$ trajectories for thermalization,
where every two adjacent configurations are separated by one auto-correlation length.

We assume that the renomalized parameters are linear in the bare parameters.
We further assume that we are in the region of light quark masses, i.e. $M^2_{PS}\sim M_q$,
and arrive at the form~\cite{AnisoL}
\beq
\xi_g(\xi^0_g,\xi^0_f,m_0)=a_0+a_1 \xi^0_g+a_2 \xi^0_f+a_3 m_0,\nn \\
\xi_f(\xi^0_g,\xi^0_f,m_0)=b_0+b_1 \xi^0_g+b_2 \xi^0_f+b_3 m_0,\nn \\
M_{PS}^2(\xi^0_g,\xi^0_f,m_0)=c_0+c_1 \xi^0_g+c_2 \xi^0_f+c_3 m_0.
\label{renormalizedpm}
\eeq

\begin{figure}[t]
\begin{center}
\includegraphics[width=.49\textwidth]{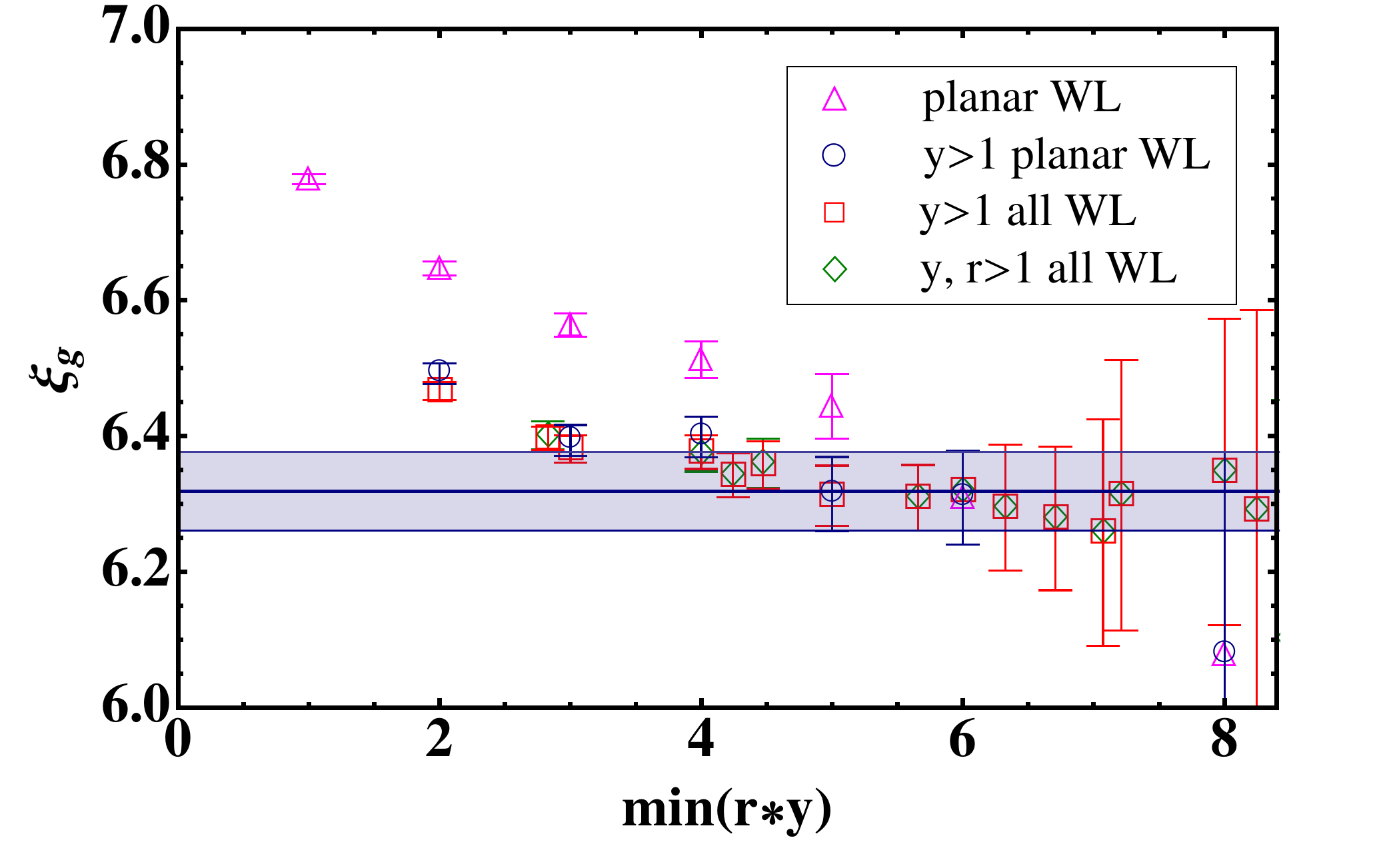}
\caption{%
Gauge anisotropy $\xi_g$ obtained using $L(\xi_g)$.
Different colors denote different sets of Wilson loops used in the calculations 
as shown in the legend. 
The blue band is the extracted value of $\xi_g$ with the statistical uncertainty. 
}
\label{xig}
\end{center}
\end{figure}

We  determine the gauge anisotropy $\xi_g$ from the static potential 
measured by Wilson loops, which was first proposed by Klassen \cite{Klassen}. 
We introduce the ratios of spatial-spatial Wilson loops, $R_s(r,y)=W_{ss}(r,y)/W_{ss}(r+1,y)$, 
and spatial-temporal Wilson loops  $R_t(r,t)=W_{st}(r,t)/W_{st}(r+1,t)$. 
The inter-quark potential at the same physical distance must have the same value, 
i.e. $R_s(r,y)=R_t(r,t=\xi_g y)$. 
In practice, we determine $\xi_g$ by minimizing $L(\xi_g)=\sum_{r,y}\ell(\xi_g;r,y)$ \cite{Umeda} with 
\beq
\ell(\xi_g)=\frac{(R_{s}(r,y)-R_{t}(r,\xi_g y))^2}{(\Delta R_s)^2+(\Delta R_t)^2},
\eeq
where $\Delta R_s$ and $\Delta R_t$ are the statistical errors of $R_s$ and $R_t$, respectively. 
In this work we extend Klassen's method by 
including nonplanar Wilson loops in which $r$ takes any two-dimentional paths in $x$-$z$ plane. 
To maximize the overlap with the physical ground state, 
we consider the path identified by $\vec{r}=(x,z)$ and $\vec{r}+1=(x+1,z)$ for fixed values of $z$. 
This modification allows us to secure enough data points 
before we encounter too large statistical errors as the size of Wilson loops increases, 
where we  find a clean signal of the convergence of $\xi_g$ to its asymptotical value. 
The numerical results in \reffig{xig} indeed show that $\xi_g$ reaches the plateau at around $\textrm{min}(r*y)=4\sim 6$. 
The largest systematic uncertainty occurs if we include the Wilson loops containing $y=1$, 
due to short-range lattice artefacts. 
In summary, the final values of $\xi_g$ is determined using planar and nonplanar 
Wilson loops, except the ones having $y=1$, at $\textrm{min}(r*y)=6$. 

We determine the fermion anisotropy $\xi_f$ from the leading-order relativistic dispersion 
relation of mesons $E^2(p^2)=M^2+p^2/\xi_f^2$ with $\vec{p}=2\pi \vec{n}/N_s$,
where $M$ is the meson mass and $\vec{n}$ is the integer vector. 
Note that the energy and mass are in units of $a_t$ while the momentum is in units of $a_s$. 
For a given momentum $\vec{p}$, we measure the energy $E(p^2)$ from a constant fit to the plateau 
of the effective mass $m_\textrm{eff}$ in the asymptotic region at a large time. 
We use point sources to construct the meson interpolating operators at source and sink, 
and consider the four lowest momentum vectors $\vec{n}=(0,0,0),\, (1,0,0),\, (0,1,0),\, (0,0,1)$ 
for the linear fit of $E^2(p^2)$, 
where the resulting fit function reproduces well the large momentum states to $|{\vec{n}}|^2=3$. 
We also find that $\xi_f$ extracted from pseudoscalar mesons 
is in good agreement with that from vector mesons and shows better precision. 
Therefore, we use the former in the tuning of lattice bare parameters. 

To determine the coefficients $a_i$, $b_i$, and $c_i$,
we perform the simultaneous $\chi^2$ fit of the numerical data to the functions in~\refeq{renormalizedpm}.
The results are
\beq
a_0=0.6(16),~a_1=0.97(13),~a_2=0.31(23),~a_3=2(4),\nn \\
b_0=1.8(24),~b_1=0.06(18),~b_2=1.1(3),~b_3=4(7),\nn \\
c_0=0.475(5),~c_1=-0.0168(4),~c_2=-0.0375(6),~c_3=0.986(11),
\label{eq:fit_results}
\eeq
where the values of $\chi^2$ per degrees of freedom are $1.72$, $0.72$, $0.23$, respectively.
These results show that the linear anzats in \refeq{renormalizedpm} 
work well over the range of considered lattice parameters. 

To determine the critical values for the bare parameters we impose the following renormalization conditions:
\beq
\xi_g(\xi^{0*}_g,\xi^{0*}_f,m^*_0)=\xi_f(\xi^{0*}_g,\xi^{0*}_f,m^*_0)=\xi,~~~
M^2_{PS}(\xi^{0*}_g,\xi^{0*}_f,m^*_0)=m^2_{ps}.
\eeq
Solving \refeq{renormalizedpm} with our target renormalized parameters of $\xi=6.3$ and $m_{ps}^2=0.005$,
we find
\beq
\xi_g^{0*}=4.84(8),~\xi_f^{0*}=4.72(12),~m_0^*=-0.2148(37).
\label{barepms}
\eeq

\subsection{Numerical results at finite temperature}
\label{finite_temperature}

\begin{figure}[t]
\begin{center}
\includegraphics[width=.48\textwidth]{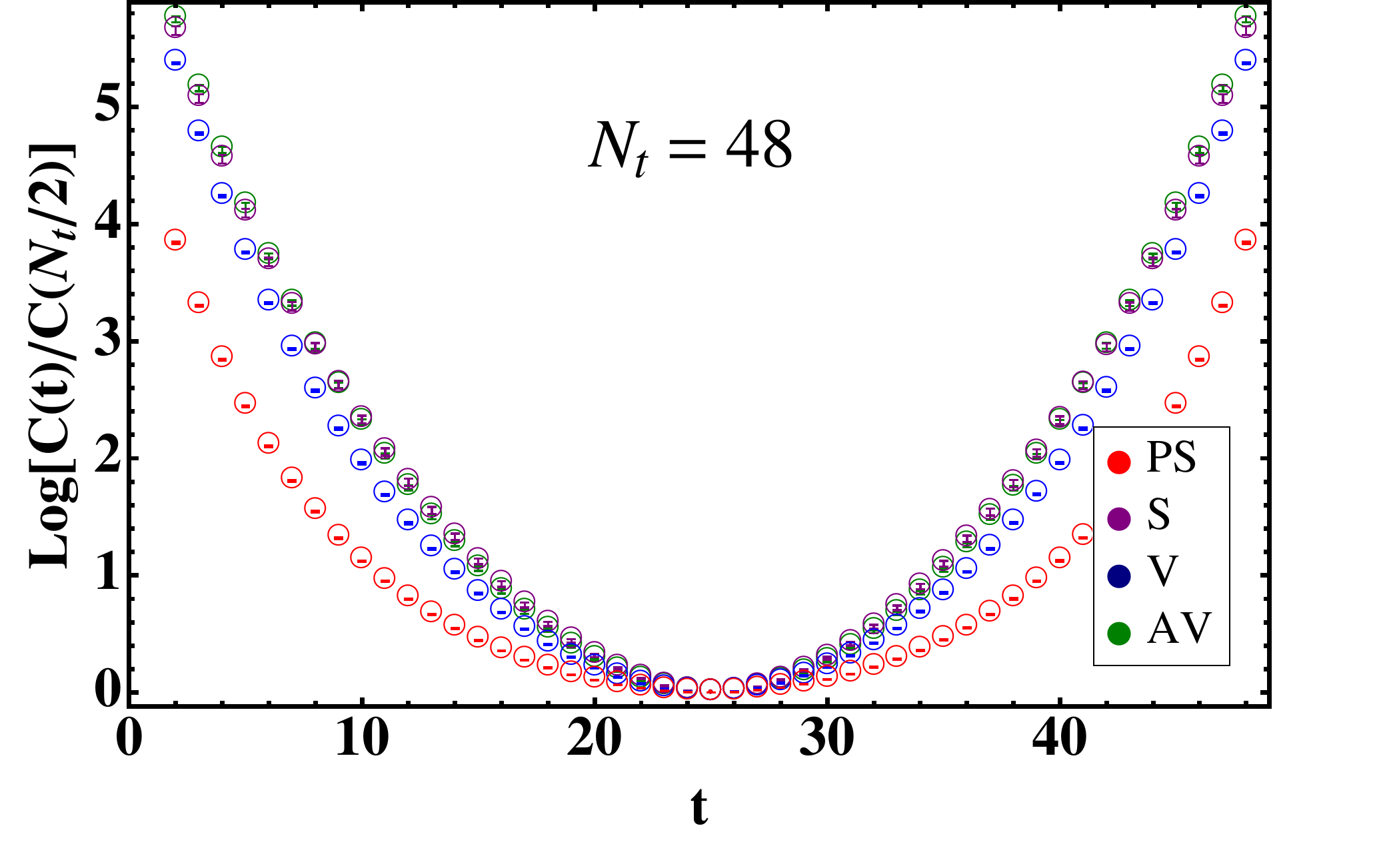}
\hskip .2in
\includegraphics[width=.48\textwidth]{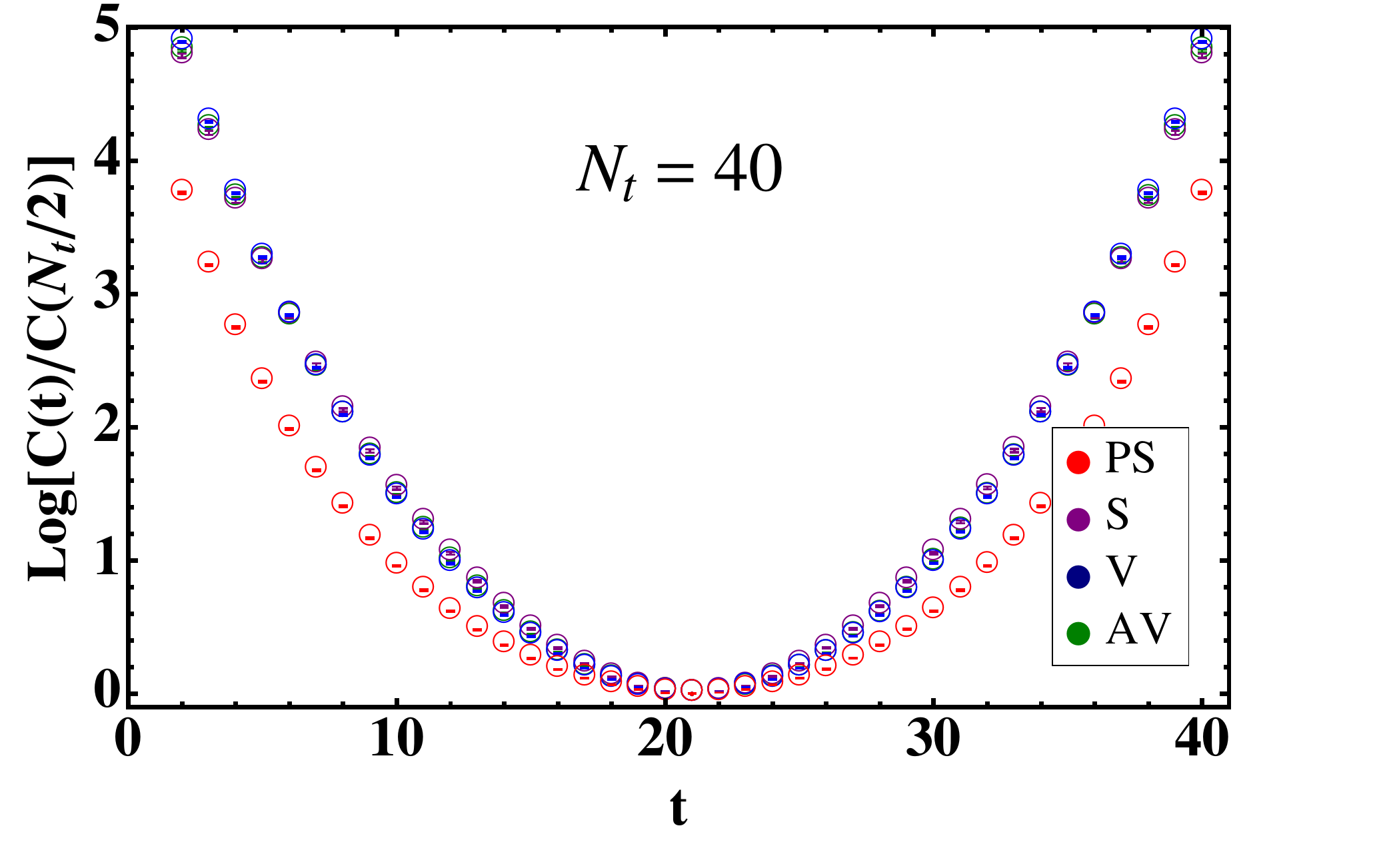}
\caption{%
Temporal correlation functions of pseudo-scalar (red), scalar (purple), vector (blue), 
and axial-vector (green) mesons, normalized by those at $N_t/2$.}
\label{tempcorr}%
\end{center}
\end{figure}

Finite $T$ calculations are performed on the anisotropic lattices of 
$N_t\times 16^3$ with $N_t=16,\,20,\,24$, $28,\,30,\,36,\,40,\,128$, 
and $N_t\times 16^2\times 24$ with $N_t=8,\,12,\,16,\,20,\,24,\,28,\,36,\,42,\,48,\,56$. 
The lattice parameters are taken from the central values in \refeq{barepms}, 
namely $\xi_g^0=4.84,~\xi_f^0=4.72,~m_0=-0.2148$.
Throughout this work, we measure $T$ in units of the pseudo-critical temperature $T_c$. 
Following the procedure described in~\cite{Aarts} , 
we determine $T_c$ using the renormalized Polyakov loop $L_R$ 
as an indication of the deconfinement crossover. 
For the determination, we consider three renormaliztion conditions, 
$L_R(N_t=24)=0.9,~L_R(N_t=24)=0.5$, and $L_R(N_t=20)=0.9$. 
From the peak of the susceptibility, $\chi(L_R)=\partial L_R/\partial T$, 
we find that $T_c a_t=0.0255(25)$ or equivalently $N_t^c=39(4)$. 
The statistical and systematic (scheme dependence) uncertainties are combined in quadrature. 

We consider flavored pseudo-scalar, scalar, vector, and axial-vector mesons,
where the corresponding interpolating fields are defined by
\beq
&\mathcal{O}_{PS}(x)=\bar{q}(x)\gamma_5 q(x),~
\mathcal{O}_{S}(x)=\bar{q}(x)q(x),&\nn \\
&\mathcal{O}_{V}^i(x)=\bar{q}(x)\gamma^i q(x),~
\mathcal{O}_{AV}^i(x)=\bar{q}(x)\gamma_5\gamma^i q(x),&
\label{mesonops}
\eeq
respectively. To improve the statistics, we use stochastic wall sources \cite{SWS} for the study
of meson spectra. 
At finite temperature the spectral function of mesons no longer exhibits a sharp peak at the mass of mesons, 
which makes it difficult to measure the mass from the single exponential analysis of a two-point function. 
In this case, it is more desirable to investigate the correlation functions by themselves. 
In \reffig{tempcorr} we show the temporal correlation functions $C(t)$ 
for $N_t=48$ and $N_t=40$, 
which exemplify the typical behaviors of $C(t)$ below and at the critical temperature, respectively.
The overlap of $C(t)$ between vector and axial-vector mesons appears at the critical temperature 
supporting the restoration of the $SU(4)$ global symmetry.

\begin{figure}[t]
\begin{center}
\includegraphics[width=.48\textwidth]{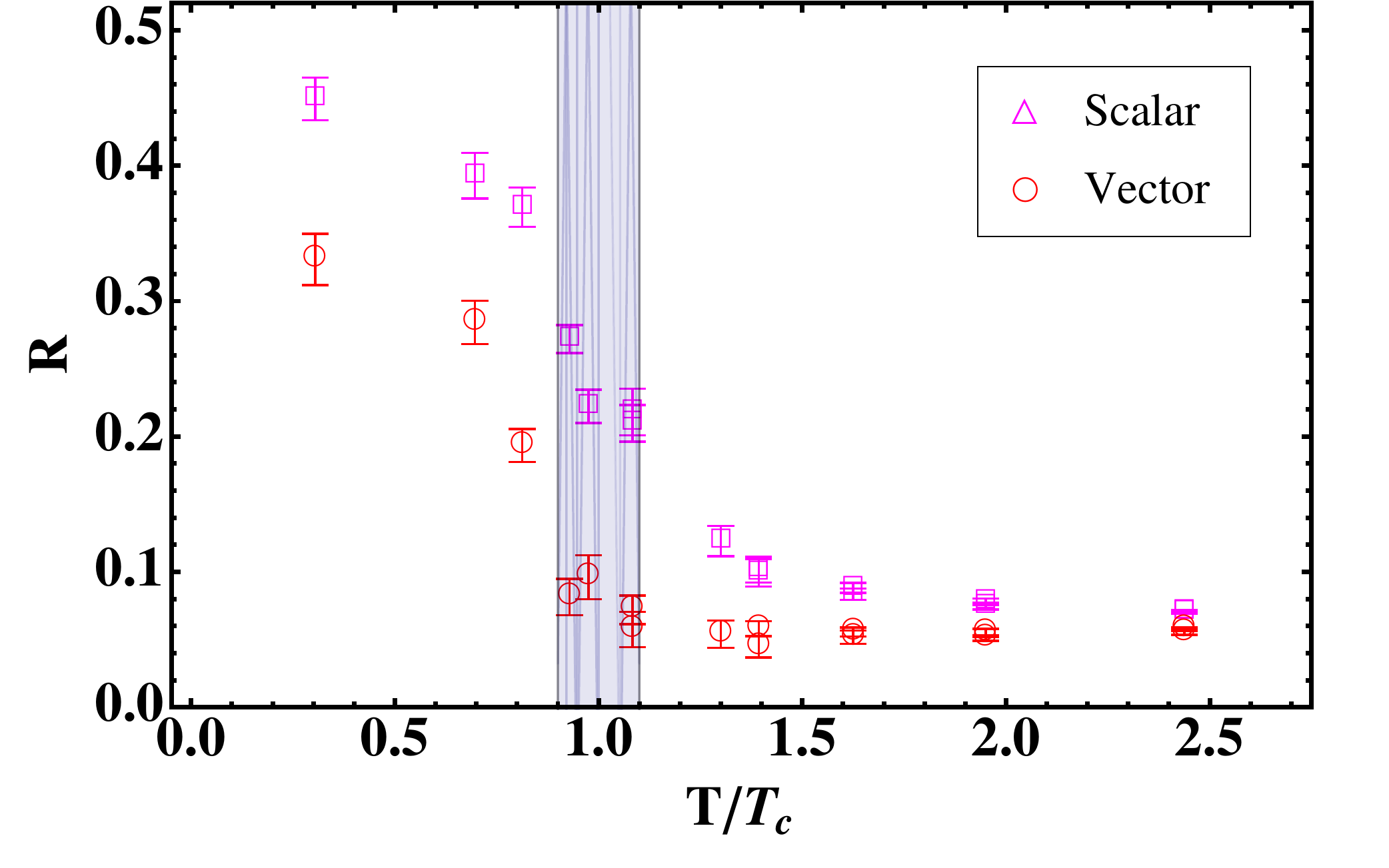}
\hskip .2in
\includegraphics[width=.48\textwidth]{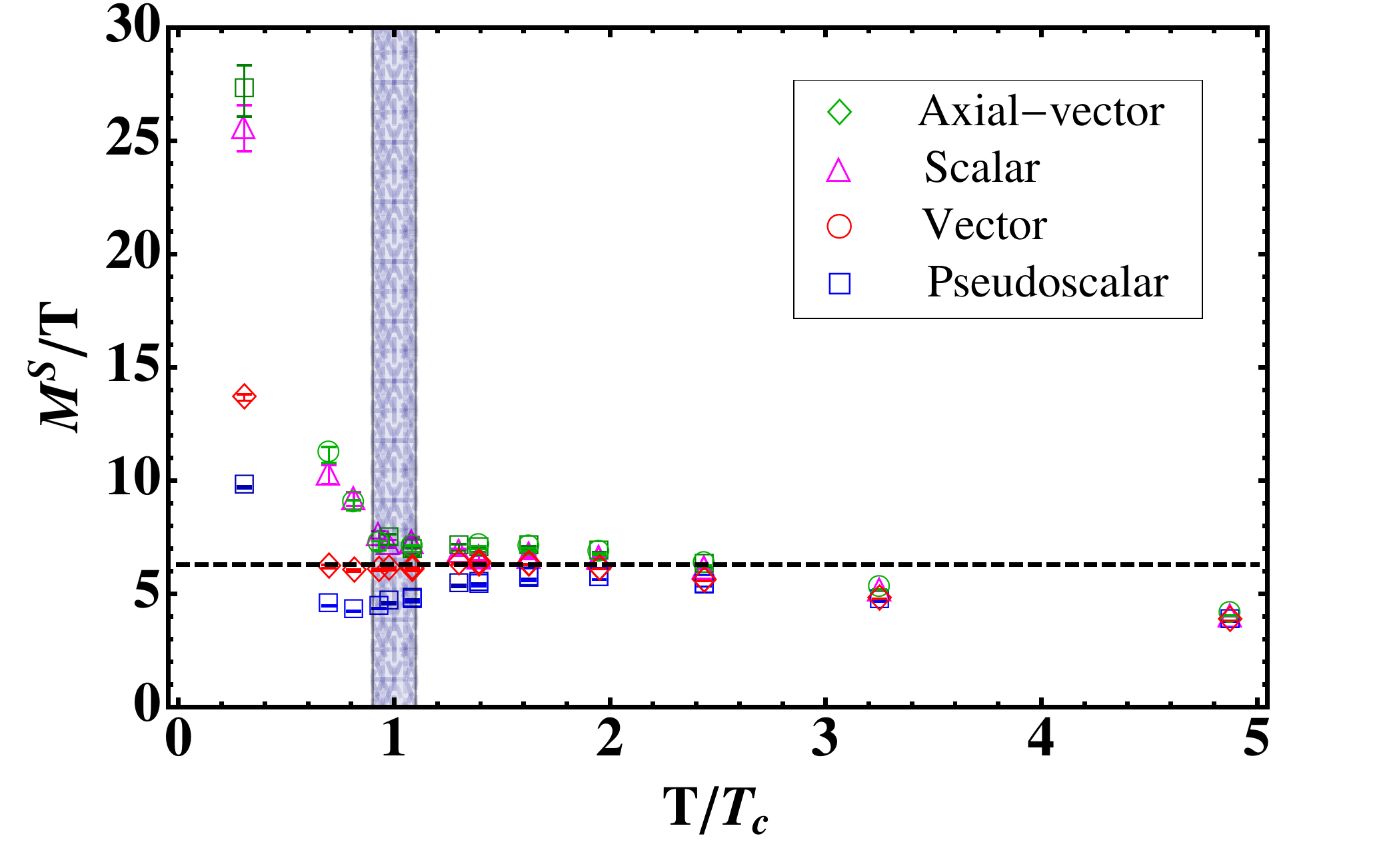}
\caption{%
Mass ratios $R$ defined in the text (left) and 
meson screening masses normalized by the temperature (right). 
The temperature is in units of  $T_c$.
The vertical blue bands denote the uncertainty of $T_c$. 
The black dashed line at $2\pi$ corresponds to the Matsubara frequency ---
the lowest excitation with free particles.  
}%
\label{smassratio}%
\end{center}
\end{figure}

In contrast to the temporal correlation function, the spatial correlation function at finite temperature
exhibits a single exponential decay at large time.
The decay rate is the {\it screening mass} that defines the effective length scale
associated with the excitation of mesonic operators in the medium \cite{DeTar}.
The screening mass is equivalent to the meson mass at zero temperature,
as the temporal and spatial correlation functions share the same spectral function. 
Using the meson interpolating fields in \refeq{mesonops}, 
we measure the screening masses from spatial correlators along $z$-direction $C(z)$. 

In addition to the screening masses, we define the mass ratios as 
\beq
R_{V(S)}(T)=\frac{M_{AV(S)}(T)-M_{V(PS)}(T)}{M_{AV(S)}(T)+M_{V(PS)}(T)}. 
\label{mass_ratio}
\eeq
These quantities are useful to quantify the size of deviation from the parity doubled states.
Our main results are presented in \reffig{smassratio}.
No significant excited state contaminations 
appear to spoil the good agreement between results from $N_z=16$ and $N_z=24$ lattices. 
Although there is small deviation from zero in the mass ratio, the plateau above $T_c$ in the vector channel
would suggest that parity partners are degenerate and thus the symmetry is effectively enhanced to $SU(4)$. 
In the scalar channel, the plateau appears at a somewhat larger temperature $\sim 1.5\,T_c$.
This result may imply that the $U(1)_A$ symmetry and $SU(4)$ symmetry are restored
at different temperature. 
This might be an indication that instead of a phase transition a crossover 
between two different regimes is taking place. 

In the finite temperature calculations, it is often suggested to plot the screening masses divided by
the temperature as it shows linear dependency above $T_c$.
The results are shown in \reffig{smassratio}, 
where those quantities approach the black dashed line at around $T_c$ and form a plateau. 
However, they start to deviate from the plateau above $2T_c$ due to the lattice artefacts (finite lattice spacing).

\section{Conclusion}
\label{Conclusion}
We considered $SU(2)$ gauge theory with $N_f=2$ flavors of fundamental Dirac fermions. 
The symmetry breaking $SU(4)\rightarrow Sp(4)$ is responsible for 
the mass splitting of flavored vector and axial-vector mesons,
while the anomalous $U(1)_A$ breaking splits scalar and pseudo-scalar mass spectra.
In the thermal bath,
one expects the restoration of the global symmetries. 
We measured the meson spectra via  
Monte-Carlo simulations on anisotropic lattices 
and indeed found  strong evidence of 
the enhancement of the global symmetry above the temperature $T_c$, 
as shown in \reffig{smassratio}. 

We only considered a single value of lattice spacing and quark mass. 
In future works we plan to extend our study
by using various choices of lattice spacings and quark masses 
to assess the size of lattice systematics. 
Another interesting direction for the future is to simulate the model with a non-zero chemical potential, 
taking advantage of the absence of the sign problem. 

\section{Acknowledgements}

This work is supported in part by the STFC Consolidated Grant ST/L000369/1.
J.-W. L. is additionally supported by Korea Research Fellowship program
funded by the Ministry of Science, ICT and Future Planning through
the National Research Foundation of Korea (2016HID3A1909283).
The authors thank S. Hands, G. Aarts, B. J{\"a}ger, F. Attanasio and E. Bennett for discussions.

\end{document}